# Testing DBMS Performance with Mutations

Xinyu Liu
Georgia Institute of Technology

Qi Zhou
Georgia Institute of Technology

Joy Arulraj
Georgia Institute of Technology

Alessandro Orso
Georgia Institute of Technology

**Abstract**

Database systems power modern data-intensive applications. It is important to ensure that they operate correctly. To this end, developers extensively test these systems to eliminate bugs that negatively affect functionality. Besides functional bugs, there is another important class of bugs that negatively affect the response time of a database system, known as performance bugs. Despite their impact on end-user experience, performance bugs have received considerably less attention than functional bugs.

In this paper, we present AMOEBA, a system for automatically detecting performance bugs in database systems. The core idea behind AMOEBA is to construct query pairs that are semantically equivalent to each other and then compare their response time on the same database system. If the queries exhibit a significant difference in their runtime performance, then the root cause is likely a performance bug in the system. We propose a novel set of mutation rules for constructing query pairs that are likely to uncover performance bugs. We introduce feedback mechanisms for improving the efficacy and computational efficiency of the tool. We evaluate AMOEBA on two widely-used DBMSs, namely PostgreSQL and CockroachDB. AMOEBA has so far discovered 39 previously-unknown performance bugs, among which developers have already confirmed 14 bugs and fixed 4 bugs.

## 1 Introduction

Database management systems (DBMSs) play a critical role in modern data-intensive applications [15, 32]. For this reason, developers extensively test these systems to improve their reliability and accuracy. They leverage tools such as SQLSMITH [4] and SQLancer [35–37] to discover crash-inducing or logic bugs in DBMSs. However, the same level of scrutiny has not been applied to *performance bugs* that increase the time taken by the DBMS to process certain queries. Delayed responses from the DBMS ultimately affect user experience [30, 42].

**CHALLENGES.** To retrieve the results for a given SQL query, the DBMS invokes a pipeline of complex components (*e.g.*, query optimizer, execution engine) [20, 33]. The overall performance of the DBMS may be reduced by sub-optimal decisions taken by any of these components and the complex interactions among them [8, 9]. Therefore performance testing on individual components of the DBMS is not sufficient [19, 22, 27, 28]. Another key challenge for detecting performance bugs in DBMS is to come up with an effective test oracle, that specifies the correct behavior (*i.e.*, response time) of a performant DBMS on a given SQL query.

There are two lines of research that attempt to address this challenge, both focusing on *performance regressions*. One approach is to use a pre-determined performance baseline as the oracle [34, 43, 44]. In particular, this technique uses a pre-commit check to ensure that the code changes do not cause an unexpected performance regression. To do so, it compares the runtime performance of the updated DBMS on a test workload against a pre-determined baseline (*e.g.*, latency < 5 seconds or throughput > 10 K transactions per second) and reports a performance bug if there is a significant deviation. While effective in detecting performance bugs that result in a regression, this approach suffers from two limitations. First, it is human-intensive and challenging to construct an accurate performance baseline and specify an error margin to account for variability in DBMS performance to reduce false positives [26]. Second, this technique relies on a fixed, limited set of queries from standard benchmarks that only cover a subset of the SQL input domain [40].

Another line of research leverages differential testing to discover performance regressions [24]. Specifically, this technique uses an oracle to compare the execution time of the same query on two versions of the DBMS and checks if the newer version takes more time to execute the query. While this technique does not require a developer-provided, pre-determined baseline, it is only able to detect *regressions*. Thus, it cannot detect a large subset of performance bugs that were always present in the system because, for these bugs, the query execution time on two versions of the DBMS would not differ. Furthermore, this technique also focuses on structurally simple queries that are tailored for uncovering regressions.

While it is also possible to detect performance bugs using metamorphic testing, it remains unclear on how to construct an effective metamorphic relation geared towards detecting performance bugs. Recently, researchers have proposed Ternary Logic Partitioning (TLP), a technique that applies metamorphic testing for detecting logic bugs in the DBMS [36]. To do this, it transforms a base query into a semantically equivalent one (called mutant query) by intentionally increasing the complexity of the base query's structure, and reports a logic bug if their result sets differ on the same database. By design, the DBMS is likely to take more time to process the mutant query in comparison to the base query. As a result, it is difficult to construct a metamorphic relation that is tailored for detecting performance bugs based on equivalent queries generated by TLP. In addition, TLP targets on comparatively simple queries that are tailored for uncovering logic bugs.

**OUR APPROACH.** In this paper, we present AMOEBA, a novel and principled tool for discovering a wider variety of performance bugs in DBMSs. AMOEBA addresses the challenges listed above along three dimensions. First, it constructs a *cross-referencing oracle* by comparing the runtime performance of *semantically equivalent queries* (*i.e.*, queries that always return the same result sets for all possible input tables) [17, 46]. When the target DBMS (only a single version is needed) exhibits a significant difference in execution time on a pair of semantically equivalent queries, the likely root cause

```sql
SELECT job, deptno FROM emp WHERE  job = 'Technical'
GROUP BY job, deptno LIMIT 13;
```
**Listing 1:** Q1. Execution Time = 13 s

```sql
SELECT CAST('Technical' AS VARCHAR(10)) AS "job", deptno
FROM emp WHERE "job" = 'Technical'
GROUP BY job, deptno LIMIT 13;
```
**Listing 2:** Q2. Execution Time = 9 ms

is a performance bug. Second, it introduces two types of semantic preserving query mutation rules that are tailored for discovering performance bugs: (1) structural mutations that transform an input query using a set of query rewrite rules (*e.g.*, join commutativity and predicate push-down) derived from the query optimization literature [21], and (2) expression mutations that modify expressions within an input query without changing the semantics of the original expressions. In this manner, AMOEBA generates semantically equivalent queries that are likely to uncover performance bugs. Third, it constructs queries tailored for discovering performance bugs, by supporting complex structures and computationally expensive SQL operators. Given the large space of SQL queries that AMOEBA can explore, we introduce a feedback mechanism that enables it to focus on exploring a subset of the query space that is more likely to uncover performance bugs.

We implement this technique in a tool called AMOEBA [11]. To demonstrate the generality of this technique, we evaluate AMOEBA on two widely-used DBMSs: CockroachDB and PostgreSQL. At the time of writing this paper, AMOEBA found 39 previously-unknown performance bugs in their query optimizers and execution engines, of which 14 bugs have been confirmed and 4 bugs have been fixed.

**CONTRIBUTIONS.** We make the following contributions:
- We present a technique for finding performance bugs using a novel application of *query equivalence* in generating a cross-referencing oracle (§2).
- We introduce two types of query mutations that preserve the semantics of queries: (1) structural mutations, and (2) expression mutations (§6).
- We present a feedback mechanism for improving the efficacy and computational efficiency of the tool (§7).
- We implemented this technique in an extensible tool called AMOEBA. In our evaluation, AMOEBA discovered 39 previously-unknown performance bugs in two widely-used DBMSs. We compare AMOEBA against two other sources of equivalent queries that could be used for detecting performance bugs: (1) a manually-written test suite in a widely-used query optimization framework, and (2) the TLP. Our empirical analysis shows how AMOEBA's generated equivalent queries are more likely to detect performance bugs, which also highlights opportunities for improving and testing future versions of these DBMSs (§8).

## 2 Motivation

In this section, we highlight the importance of detecting performance bugs in DBMSs using a motivating example.

**EXAMPLE.** Listing 1 and Listing 2 show a pair of *semantically equivalent* queries based on the SCOTT schema [10]: Q1 and Q2. Although Q1 and Q2 are equivalent, Q1 runs 1444× slower than Q2 on the same database in CockroachDB [6] (v20.2.0-alpha).

With the first query, the DBMS picks the vectorized engine to fetch tuples, which applies the UNORDERED DISTINCT operation on the entire *emp* table. On the other hand, with the second query, the DBMS chooses the row engine, which executes the UNORDERED DISTINCT operation on the *emp* table only after taking the LIMIT operator into consideration. As a result, while Q1 takes 13 seconds to execute on a table that contains 10 million rows, Q2 only takes 9 milliseconds. The DBMS developers have acknowledged that this is a previously-unknown performance bug that stems from a limitation in the execution engine. In particular, the vectorized execution engine does not propagate the hint from the LIMIT operator to the other operators (*i.e.*, HASH JOIN and UNORDERED DITINCT). Since these buffering operators are computationally expensive and critical to query performance, the CockroachDB developers quickly fixed this performance bug in the UNORDERED DISTINCT operator within a week and plan to fix the bug in the HASH JOIN operator in the future. This example highlights the existence of performance bugs in DBMSs and their significant impact on query performance.

## 3 Background

To better appreciate the internals of AMOEBA, we now provide a brief overview of semantically-equivalent queries.

**SEMANTIC EQUIVALENCE.** Two queries *Q1* and *Q2* are semantically equivalent if they always return the same results on any input database instance. Query equivalence is a well-studied topic and is used in many applications: (1) testing correctness of DBMS and SQL queries [36, 38], (2) educating developers [25], and (3) automatically grading student assignments [14]. Unlike prior efforts, we seek to leverage semantic equivalence of queries to find performance bugs in DBMSs.

**QUERY REWRITING.** AMOEBA constructs equivalent queries by rewriting them using a set of rules that preserve equivalence [21]. A representative rule consists of using values from filters to mutate projection columns. Consider the query shown in Listing 1. For this query, we illustrate how this rule transforms its projection columns while preserving semantic equivalence in Figure 1.

The logical query plan of Q1 is shown in Figure 1a. Since the filter clause selects tuples wherein the *job* attribute equals to a specific string value, the rule replaces the final projection column *job* with a literal column that takes the same string value. Figure 1b shows the transformation result, which represents the logical query plan of Q2.

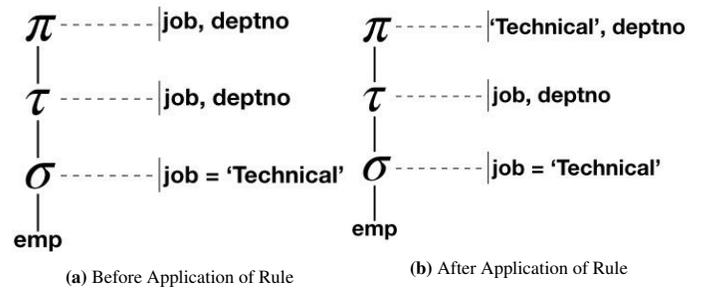

(a) Before Application of Rule  (b) After Application of Rule

**Figure 1: Query Rewriting –** The impact of the projection column mutation rule on the query shown in Listing 1.

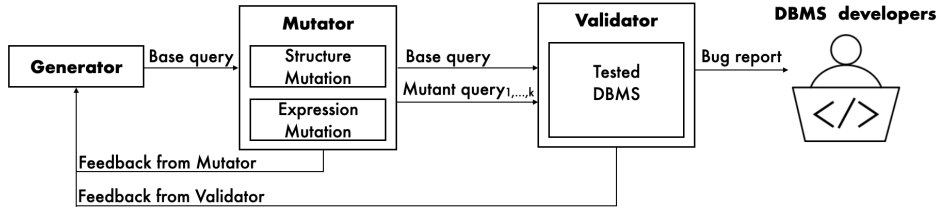

**Figure 2: Architecture of AMOEBA** – GENERATOR constructs a set of base SQL queries based on feedback from other components. MUTATOR performs semantic-preserving structural and expression mutations on the base queries. VALIDATOR executes a pair of semantically equivalent queries using the target DBMS and reports query pairs that exhibit a significant difference in runtime performance.

## 4 System Overview

AMOEBA help developers uncover performance bugs in a DBMS. The key idea is to compare the runtime performance of semantically equivalent queries on the same DBMS. Specifically, we expect the DBMS to execute equivalent queries in the same amount of time. If that is not the case and the difference in the query execution time exceeds a developer-specified threshold (*e.g.*, 2×), then AMOEBA has found a performance bug. Such a *cross-referencing oracle* allows us to detect performance bugs in a single DBMS (*i.e.*, without resorting to comparative analysis against another DBMS).

In this section, we present an overview of the key components of AMOEBA. Figure 2 illustrates the architecture of AMOEBA. AMOEBA contains three components: (1) GENERATOR, (2) MUTATOR, and (3) VALIDATOR.

❶ GENERATOR leverages a domain-specific fuzzing technique to generate SQL queries. We refer to these queries as *base queries*. GENERATOR is tailored to generate queries that are more likely to trigger performance bugs in DBMSs. In particular, it receives feedback from the latter components of AMOEBA to guide the query generation process.

❷ MUTATOR takes a base query as input and seeks to generate equivalent queries. To do so, it applies a set of semantics-preserving query rewriting rules on the given query. We refer to the resulting set of equivalent queries as *mutant queries*. The output of this component is, thus, the base query and a set of equivalent mutant queries.

❸ VALIDATOR takes a set of equivalent queries as input and generates a list of performance bug reports. VALIDATOR runs each pair of equivalent queries on the target DBMS and observes whether any pair exhibits a significant difference in their runtime performance. If that is the case, it first verifies if this behavior is reproducible across multiple runs. After confirming the bug, it generates a report that consists of: (1) a pair of equivalent queries that exhibit different runtime behavior, and (2) their query execution plans. We manually submit these reports to the DBMS developers for their consideration.

We next present the three components of AMOEBA in §5, §6, and §7, respectively.

## 5 Query Generator

AMOEBA uses a grammar-aware GENERATOR that randomly constructs a set of base queries. As shown in Algorithm 1, GENERATOR takes a target database as input and generates base queries as output. At a high level, GENERATOR uses two key procedures for generating

**Algorithm 1:** Algorithm for generating SQL queries

**Input** : *database*: database under test
**Output**: *base queries*: random SQL queries

1. meta-data ← RetrieveMetaData(*database*);
2. prob_table ← InitProbabilityTable();
3. rule_activation_frequency_table ← InitRuleActivationFrequencyTable();
4. **Procedure** GenerateQuery(*meta-data, prob_table*)
5.    **while** *True* **do**
6.       specification ← BuildSpecification(*meta-data, prob_table*) ;
7.       base query ← SpectoQuery(*specification, dialect*) ;
8.       **return** *base query* ;
9.    prob_table ← UpdateProbTablewithFeedback(*base query, prob_table*) ;

10.

11. **Procedure** UpdateProbTablewithFeedback(*base query, prob_table*)
12.    **if** MutateQuery(*base query*) **is** True **then**
      // If the base query can be mutated to generate equivalent queries, update the probability table so that it is more likely to generate queries containing SQL entities that the base query contains.
13.       UpdateProbTableWithMutatorFeedback(*base query, prob_table, applied_rules*) ;
14.    **if** TriggerPerformanceBug(*base query*) **is** True **then**
      // If the base query can trigger a performance bug, update the probability table so that it is more likely to generate queries containing SQL entities that the base query contains.
15.       UpdateProbTableWithValidatorFeedback(*base query, prob_table*) ;

16.

**Table 1: SQL Grammar –** A subset of the SQL grammar that allows to generate queries with a variety of structures and operators.

| | | |
|---|---|---|
| query_spec | ::= | SELECT <column$_{ref}$> FROM <table$_{ref}$> <group_clause> <limit_clause> |
| table$_{ref}$ | ::= | <table$_{simple}$> \| <table$_{joined}$> |
| table$_{joined}$ | ::= | <table$_{ref}$> <join_spec> <table$_{ref}$> |
| join_spec | ::= | <join$_{type}$> ON <join$_{cond}$> |
| join$_{type}$ | ::= | LEFT \| CROSS \| INNER |
| join$_{cond}$ | ::= | <bool$_{expr}$> \| TRUE |

**Table 2: Probability Tables –** A list of probability values that are used to generate table references and join conditions.

| (a) Table References | | | | (b) Join Conditions | |
|---|---|---|---|---|---|
| | | table$_{joined}$ | | | |
| table$_{simple}$ | LEFT | CROSS | INNER | bool$_{expr}$ | TRUE |
| 0.5 | 0.16 | 0.17 | 0.17 | 0.5 | 0.5 |

queries that are more likely to trigger performance bugs: (1) GENERATEQUERY (§5.1) uses a top-down, grammar-aware approach to generate queries that are compatible with the input database's schema, and (2) UPDATEPROBTABLEWITHFEEDBACK (§5.2) leverages feedback from prior runs of MUTATOR (§6) and VALIDATOR (§7) to guide the GENERATEQUERY procedure. AMOEBA relies on this feedback mechanism to improve the probability of generating queries that trigger performance bugs. Next, we explain these two procedures in detail.

## 5.1 Grammar-Aware Query Generation

Researchers have extensively explored techniques for grammar-aware query generation [4, 12, 13, 45]. AMOEBA's query generation approach differs from prior systems in that it is geared towards generating queries that are *more likely* to trigger performance bugs in DBMS. To accomplish this, AMOEBA contains two components: (1) a *grammar* for generating queries with different structures and operators, and (2) a *probability table* defined with respect to the grammar to guide the query generation process. We next describe these components.

GRAMMAR. AMOEBA uses a grammar based on the SQL-92 standard [1]. The grammar is expressed in Backus–Naur Form (BNF), which consists of both terminal and non-terminal symbols. Table 1 lists a subset of the grammar that allows AMOEBA to generate queries with a variety of structures and operators. For instance, the non-terminal symbol $table_{ref}$ may either be a base table (*i.e.*, $table_{simple}$) from the target database or a derived table (*i.e.*, $table_{joined}$) resulting from a JOIN operator.

PROBABILITY TABLE. As shown in Table 2, AMOEBA maintains a probability table to track the likelihood of using each non-terminal and terminal symbol when generating queries following the grammar. Thus, this table guides the likelihood of a SQL structure or clause appearing in the generated query. For all symbols that stem from a given non-terminal symbol, the probabilities sum up to one. For instance, Table 2b specifies that there is an equal chance to generate a non-terminal symbol, the JOIN condition, using either a boolean expression (*e.g.*, t1.k = t2.k) or the keyword TRUE.

Next, we walk through the algorithm of GENERATEQUERY. The first step is to acquire information needed for generating queries (line 1): (1) it randomly samples a small dataset from the target database. This dataset will later be used to generate predicates in queries, which allows to generate predicates with a variety of selectivity (requires data from the database content). (2) it also collects table schemas of the target database (*i.e.*, table and column names, and column types). GENERATOR later will rely on such information to create valid expressions, such as SQL function calls and column comparison (both require taking types into consideration). Secondly, it initializes a probability table for guiding the query generation procedure (line 2), wherein each entry has a default value. Next, the procedure invokes the BUILDSPECIFICATION function to construct a query specification based on: (1) the SQL grammar, and (2) the collected meta-data (line 6). Lastly, GENERATOR translates the specification into a well-formed query for the target DBMS (line 7).

## 5.2 Feedback from Mutator and Validator

We now discuss how GENERATOR updates the probability table based on the feedback from MUTATOR and VALIDATOR (line 9).

FEEDBACK FROM MUTATOR. GENERATOR uses the feedback from the MUTATOR to improve the likelihood of generating base queries that can be successfully transformed by MUTATOR. Since AMOEBA relies on the generation of semantically equivalent queries, this feedback mechanism indirectly increases the likelihood of discovering performance bugs. GENERATOR updates the probability table when a base query that it generates is successfully transformed by MUTATOR (line 12). UPDATEPROBTABLEWITHMUTATORFEEDBACK is the relevant procedure (line 13), which contains two steps. First, it extracts SQL entities from the base query. For example, since Q1 in Listing 1 can be successfully mutated into Q2, GENERATOR extracts the following entities from Q1: $table_{simple}$, GROUP BY and LIMIT. This design decision is based on the assumption that these entities are correlated with successful mutations. Then, GENERATOR updates the probability table by increasing values of these extracted entities and decreasing values of the other entities. For example, given Q1's extracted entities, it updates the probability table as follows: it decreases the probabilities of LEFT, CROSS, and INNER, and increases the probability of $table_{simple}$, GROUP BY and LIMIT. After this update, GENERATOR will generate base queries that are more likely to contain the GROUP BY and LIMIT clauses.

GENERATOR also seeks to avoid generating queries that only trigger a limited set of mutation rules. This improves the computational efficiency of AMOEBA (§8.4). To accomplish this, it keeps track of the frequency with which each mutation rule has been fired (line 3). For instance, when a base query triggers a less-frequently triggered mutation rule, GENERATOR bumps up the probabilities of the query's extracted entities with a larger increment. In this way, GENERATOR constructs queries that trigger a wide variety of mutation rules.

FEEDBACK FROM VALIDATOR. GENERATOR utilizes information about discovered performance bugs to increase the likelihood of generating base queries that uncover them. When VALIDATOR discovers a base query that eventually leads to finding a performance bug (line 14), GENERATOR invokes the UPDATEPROBTABLEWITHVALIDATORFEEDBACK procedure to update the probability table accordingly (line 15). Similar to UPDATEPROBTABLEWITHMUTATORFEEDBACK, this procedure first extracts the SQL entities from the base query and then updates the probability table accordingly. In this way, GENERATOR constructs queries that are more likely to uncover performance bugs.

## 6 Query Mutator

MUTATOR is the next key component of AMOEBA. It takes a base query as input and seeks to generate mutant queries that are semantically equivalent to the base query.

QUERY EQUIVALENCE. While the query equivalence property has many applications [14, 25, 38], it has never been utilized for uncovering performance bugs. Notably, SQLancer is a system that constructs equivalent queries for discovering *logic bugs* in DBMSs. In particular, it generates equivalent queries using TLP [36]. We defer a comparative analysis of AMOEBA against SQLancer to §8.6.

At a high level, MUTATOR is a framework for rewriting queries using a set of semantics-preserving query transformation rules. We next present the internals of MUTATOR with a description of its mutation rules in §6.1 and its algorithm in §6.2.

Table 3: List of Illustrative Mutation Rules.

| Category | Rule Idx | Transformation | Working Example |
|---|---|---|---|
| Structure | 0 | Push aggregate function through JOIN | t1 JOIN t2 GROUP BY t1.c → (t1 GROUP BY t1.c) JOIN t2 |
|  | 13 | Push filter through GROUP BY | (t1.c GROUP BY t1.c) WHERE t1.c > 0 → (t1.c WHERE t1.c > 0) GROUP BY t1.C |
|  | 16 | Push filter expression through JOIN | t1 JOIN t2 WHERE t1.c = 5 → (t1 where t1.c = 5) JOIN t2 |
|  | 51 | Remove SORT on empty table | (t1 WHERE false) ORDER BY t1.c → t1 WHERE false |
|  | 59 | Propagate LIMIT through LEFT JOIN | t1 LEFT JOIN t2 LIMIT 5 → (t1 LIMIT 5) LEFT JOIN t2 LIMIT 5 |
| Expression | 15 | Convert EXTRACT into date range comparison | EXTRACT (YEAR FROM c1) < 2021 → c1 < TIMESTAMP '2021-01-01 00:00:00' |
|  | 54 | Reduce expression in FILTER | WHERE c1 = CAST(10/2) as INT → WHERE c1 = 5 |
|  | 55 | Reduce expression in JOIN condition | t1 JOIN t2 ON CAST (t1.integer_primary_key) as INT = t2.integer_primary_key |

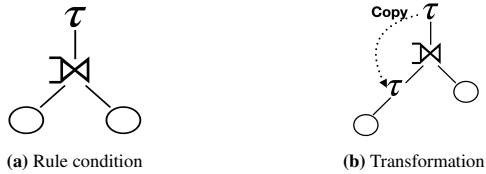

(a) Rule condition     (b) Transformation

Figure 3: Illustrative Rule Condition and Transformation (Rule 59).

## 6.1 Mutation Rules

MUTATOR currently supports 75 mutation rules that fall into two categories: (1) rules for mutating query structure (67 rules), (2) rules for mutating SQL expressions (8 rules). Table 3 presents a subset of these rules to illustrate their variety. A complete list of rules is available at [11].

Each rule is represented by a triple: <*id, condition, transformation*> [21]. Specifically, to guarantee that a given rule does not modify the semantics of the input query, MUTATOR checks whether the input query satisfies the rule's *condition*. If so, it applies the *transformation* on the input query to generate a semantically equivalent query. We next present the two types of rules.

**RULES FOR MUTATING QUERY STRUCTURE.** These rules modify the query's structure based on a relational algebraic transformation that preserves equivalence. In order to preserve semantics, each rule performs transformation on the input query only if a specific rule condition is met. Figure 3 illustrates the condition and transformation associated with rule 59. It modifies the query structure by pushing the LIMIT operator below the LEFT JOIN operator. As shown in Figure 3a, this rule is only triggered when there is a LIMIT operator on top of a LEFT JOIN operator. If the input query satisfies this condition, then this rule transforms it to a structurally-different query shown in Figure 3b by pushing a copy of the LIMIT operator and its argument below the LEFT JOIN operator.

**RULES FOR MUTATING EXPRESSION.** These rules rewrite the expression in the query without altering the query structure. Consider rule 15 that mutates the comparison between a constant value with the result of the EXTRACT function (*e.g.*, EXTRACT (YEAR FROM c1)) < 2021). To do this, the rule replaces the original expression with a new one that compares the attribute against a timestamp.

**COMPOSITIONAL EFFECTS.** Since mutation rules are sequentially applied on the input query, their compositional effects further expand the query space explored by MUTATOR [31]. We now present an example that illustrates how the order in which the rules are applied determines the generated mutant query.

```
SELECT  *
FROM    (t1 where EXTRACT (YEAR FROM t1.c1) = 2021)
JOIN    t2
WHERE   EXTRACT (YEAR FROM t1.c1) > EXTRACT (YEAR FROM
        t2.c2)
```
Listing 3: Q5. Compositional Effects

Consider the query shown in Listing 3 that performs a JOIN of tables t1 and t2 with two filter predicates invoking the EXTRACT functions: one filter predicate is before the JOIN and the other one is on top of the JOIN results. Consider the following two rules that are applied sequentially on the input query: (1) rule 15 that reduces the EXTRACT function, and (2) rule 55 that reduces the expression in JOIN condition by replacing variable with known constants. If MUTATOR applies rule 15 before rule 55 on the input query, it only reduces the EXTRACT function in the lower predicate and creates a new comparison between t2.c2 and a timestamp value. Then the pattern of the lower filter predicate no longer satisfies the condition for exercising rule 55. The resulting query's logical plan is shown in Figure 4a. On the other hand, if MUTATOR applies rule 55 before rule 15, both rules will be exercised and leads to a different mutant query. Specifically, rule 55 first transforms the upper predicate into an equivalent one: $2021 > EXTRACT (YEAR FROM t1.c1)$. Then rule 15 reduces EXTRACT functions in both predicates. The resulting logical plan is shown in Figure 4b.

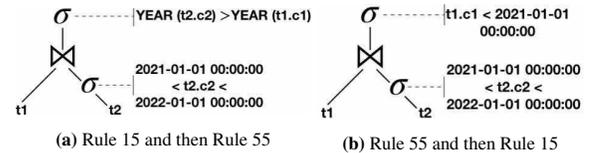

(a) Rule 15 and then Rule 55     (b) Rule 55 and then Rule 15

Figure 4: Compositional Effects – The impact of the order of application of mutation rules on the query shown in Listing 3.

## 6.2 Mutation Algorithm

We outline the algorithm of MUTATOR in Algorithm 2. MUTATOR takes a base query and metadata of the target database as input and returns the base query and its semantically equivalent mutant queries as output. It first preprocesses the base query and generates its logical query plan tree $R_{origin}$ on which it applies the mutation rules (line 2). It attempts to mutate $R_{origin}$ for a total of *number_of_-attempts* times (line 5).

**Algorithm 2:** Procedure for mutating SQL queries

```
Input  : base query; meta-data: metadata for the target database
Output : base query and mutant queries
1  Procedure MutateQuery(base query, meta-data)
2      R_origin ← Preprocess(base query) ;
3      transformed_trees ← EmptySet() ;
4      mutant queries ← EmptySet() ;
5      for k ← 1 to number_of_attempts do
              // randomly select a list of mutation rules
6          mutate_rules ← RulesInitialization() ;
7          R_new ← MutateTree(R_origin, mutate_rules, meta_data) ;
8          if R_new ∉ transformed_trees then
9              new query ← TranslateToQuery(R_new, dialect);
10             Update(transformed_trees, mutant queries) ;
11     return base query, mutant queries ;
12 Procedure MutateTree(R_origin, mutate_rules, meta_data)
13     target_expr ← R_origin;
14     for rule ∈ rewrite_rules do
15         target_expr ← ApplyRule(target_expr, rule, meta_data) ;
16     if target_expr ≠ R_origin then
17         return target_expr ;
```

In each iteration, the MUTATOR performs the following three steps. ❶ It randomly initializes an ordered list of mutation rules (*mutate_rules*) that it then applies on the $R_{origin}$ in a sequential manner (line 6). In doing so, MUTATOR increases the likelihood of uncovering different compositional effects of mutation rules on the input query. ❷ Next, it invokes the MUTATETREE procedure to transform $R_{origin}$ using *mutate_rules* (line 7). Within this procedure, MUTATOR uses the *meta-data* to decide whether the *rule condition* for performing the transformation is met (line 15). The MUTATETREE procedure returns the resulting plan tree $R_{new}$ only if $R_{new}$ is different than the $R_{origin}$ (line 17). ❸ After getting $R_{new}$ from the MUTATETREE procedure, MUTATOR checks whether it is different from trees constructed in prior mutation attempts (line 8). If so, MUTATOR translates back $R_{new}$ into a well-formed SQL query (*new query*) based on the target DBMS's dialect and appends it to mutant queries (line 9, line 10). Finally, it returns the base query and mutant queries as its output.

## 7 Validator

VALIDATOR is the last component of AMOEBA. It takes a set of semantically-equivalent queries as input and generates performance bug reports as output. In particular, VALIDATOR compares the execution time of these queries as a *cross-referencing oracle* for detecting performance bugs in DBMS. If a pair of equivalent queries consistently exhibit significant difference in their runtime performance, it generates a performance bug report that consists of: (1) the pair of queries, and (2) their execution plans. Before presenting the algorithm of VALIDATOR, we discuss two challenges associated with discovering performance bugs based on query equivalence.

**VARIATION IN EXECUTION PLANS.** A pair of semantically equivalent queries with different syntax (*i.e.*, structural difference or predicate difference) may reduce to the same query execution plan. In this case, the DBMS will execute these queries in the same way and there will not be any difference in runtime performance. Such queries are not *useful* for discovering performance bugs. To improve the computational efficiency of AMOEBA, VALIDATOR focuses on equivalent queries that have different execution plans. In particular, before executing a set of equivalent queries and comparing their runtime performances, it first compares their plans and *skips* this query pair if they have the same plan.

**FILTERING OUT FALSE POSITIVES.** The runtime performance of a query may be affected by system-level factors (*e.g.*, caching behavior of concurrent queries) [22]. To avoid false positives stemming from such factors, before reporting a discovered performance bug to the developers, VALIDATOR verifies that the difference is consistently reproducible by re-executing the same query pair multiple times in isolation and in different execution orders.

**Algorithm 3:** Procedure for detecting performance bugs

```
Input  : base query                    mutant queries
         validation_attempt: # of additional runs    threshold: least time difference
Output : bug reports: each report contains a query pair and execution plans
1  if CheckPlanDiff(base query, mutant queries) then
       // Run equivalent queries if they have different execution plans
2      time_list ← RunQuery(base query, mutant queries) ;
3      if Max(time_list) > threshold ∗ Min(time_list) then
           // Rerun queries for validation_attempt times to confirm the
              difference is consistent
4          if Confirm(base query, mutant queries) then
5              GenBugReport(base query, mutant queries) ;
6
7  Procedure CheckPlanDiff(base query, mutant queries)
8      cost_list ← EstimateCost(base query, mutant queries) ;
9      if Count(Set(cost_list)) > 1 then
10         return True ;
11
12 Procedure Confirm(base query, mutant queries, validation_attempts)
13     difference_count ← 1 ;
14     for k ← 1 to validation_attempts do
15         time_list ← RunQuery(base query, mutant queries) ;
16         if Max(time_list) > threshold ∗ Min(time_list) then
17             difference_count += 1 ;
18     if difference_count = validation_attempts then
19         return True ;
20
```

Algorithm 3 presents the algorithm of VALIDATOR. ❶ It first invokes the CHECKPLANDIFF procedure to filter out equivalent queries that lead to equivalent execution plans (line 1). VALIDATOR assumes that two queries have the same execution plan if their estimated costs are the same. To compare estimated costs for query plans, VALIDATOR utilizes the EXPLAIN feature of DBMSs (line 8) [39]. If VALIDATOR determines that the estimated plan costs are identical, it assumes that the query set converges to the same plan and skips them (line 9). ❷ After getting equivalent queries with different plans, VALIDATOR runs them on the DBMS and records their execution time (line 2). Then, within the resulting set of execution times, VALIDATOR checks whether the ratio of the longest to the shortest query execution time exceeds a developer-specified threshold (line 3). If the ratio exceeds the threshold, VALIDATOR invokes the CONFIRM procedure to check whether the runtime performance difference is consistently reproducible (line 4). ❸ Within the CONFIRM procedure, VALIDATOR re-executes these queries on the DBMS for multiple runs in random orders and monitors whether the execution time difference still holds (line 14 – line 19). If the difference is consistent, then VALIDATOR automatically generates a performance bug report (line 5).

# 8 Evaluation

To evaluate the effectiveness and generality of AMOEBA, we investigate the following questions:

**RQ1.** Can AMOEBA find performance bugs in DBMSs? (§8.3)
**RQ2.** What is the efficiency of AMOEBA? (§8.4)
**RQ3.** Are all mutation rules created equal with respect to discovering performance bugs? (§8.5)
**RQ4.** How does AMOEBA compare against other techniques for finding performance bugs? (§8.6)
**RQ5.** How do the base queries in AMOEBA compare against those in Calcite? (§8.7)

## 8.1 Implementation

We now discuss the implementation details of AMOEBA. AMOEBA currently supports two widely-used DBMSs: (1) CockroachDB, and (2) PostgreSQL.

**QUERY GENERATION.** We design a GENERATOR that constructs statements that are more likely to be syntactically correct and that covers a widely-supported subset of SQL constructs [1]. To this end, we design a GENERATOR based on SQLALCHEMY [7], a SQL toolkit and Object Relational Mapper (ORM) written in Python. The design and implementation of the GENERATOR is beyond the scope of this paper.

**QUERY MUTATION.** We build MUTATOR on top of the Calcite [5] query optimization framework. Calcite transforms queries by iteratively applying a set of query rewrite rules [21]. Calcite works well with SQLALCHEMY in that they both cover widely-supported subset of SQL constructs and dialects. So, most of the semantically-valid queries constructed by GENERATOR are successfully parsed by the MUTATOR. AMOEBA is extensible by design since it leverages the SQLALCHEMY and Calcite frameworks.

**IMPLEMENTATION SCOPE.** AMOEBA supports queries with four data types: integer, double, datetime, and string types. Additionally, the queries may use several SQL constructs (*e.g.*, GROUP, DISTINCT, ORDER and UNION) and functions (*e.g.*, AVG and SUM). We present a detailed list of supported SQL constructs in [11].

**BENCHMARK.** We use a database derived from the SCOTT schema [10] in our experiments. This is because we seek to compare AMOEBA against the manually-crafted Calcite test suite that is based on the SCOTT schema. The SCOTT schema is comparatively simple: only contains three tables with two primary keys and one foreign key. We configure the size of the SCOTT-based database to 30 MB. The query execution time is proportional to the size of the database. We adjust the database size to achieve a balance between discovering reproducible bugs (requires a larger database) and improving the computational efficiency of AMOEBA (requires a smaller database). In particular, we ensure that most queries finish within the developer-specified, time-out period for a query (*i.e.*, 15 seconds). To support larger databases and more complex queries, we must increase this time-out period. We configured this temporal constraint based on the feedback received from database developers.

## 8.2 Evaluation Setup

We focus on two DBMSs: (1)CockroachDB (v20.2.0-alpha), and (2)PostgreSQL (v12.3) . We run all experiments on a server with two Intel(R) Xeon(R) E5649 CPUs (24 processors) and 236 GB RAM. We manually examine the bug reports generated by AMOEBA and report them to developers for their feedback.

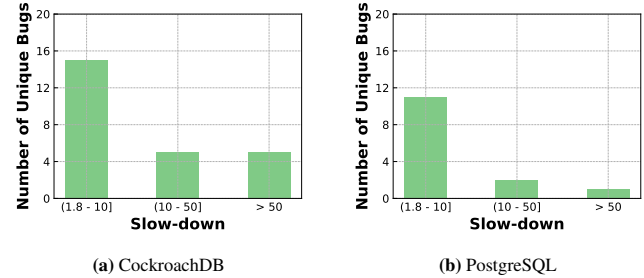

(a) CockroachDB  (b) PostgreSQL

**Figure 5: Runtime Impact of Discovered Performance Bugs** – Figures (a) and (b) show the slow-down of queries caused by performance bugs for CockroachDB and PostgreSQL, respectively.

## 8.3 RQ1 — Performance bugs detection

AMOEBA found 39 previously-unknown performance bugs. Figure 5 summarizes the impact of the discovered performance bugs (*i.e.*, the performance gap between equivalent query pairs in the bug report).

**RUNTIME IMPACT.** While bugs found in CockroachDB exhibit a slow-down ranging from 1.9× to 669.1×, those found in PostgreSQL exhibit a slow-down ranging from 1.9× to 555.6× for equivalent queries. Developers have confirmed 14 bugs and fixed 4 of them in the `master` branch of these DBMSs.

**DESCRIPTION OF BUGS.** We now discuss a subset of performance bugs found by AMOEBA to illustrate the types of bugs it finds. A complete list of reported bugs is provided in [11].

**EXAMPLE 6. EXPRESSION SIMPLIFICATION.** The pair of equivalent queries below exhibit a 3.2× slow-down in CockroachDB. This performance bug is present in the query optimizer. While both queries request the DBMS to fetch the same set of tuples, the DBMS chooses to use different execution plans to process them.

```sql
/* [First query, 75 milliseconds] */
SELECT Max(emp.sal)
FROM   dept INNER JOIN emp ON ename NOT LIKE name
WHERE  name = 'ACCT';

/* [Second query, 238 milliseconds] */
SELECT Max(emp.sal)
FROM   dept INNER JOIN emp ON ename NOT LIKE name
WHERE  name = 'ACCT' IS TRUE;
```

With the first query, the DBMS propagates the value of *name* from the predicate into the JOIN condition. Accordingly, it scans a subset of the table *emp* using a FILTER derived from the JOIN condition (*i.e.*, ename NOT LIKE 'ACCT'). On the other hand, with the second query, the DBMS chooses to scan the entire table *emp*, resulting in a more expensive HASH JOIN than the first query. After analyzing this bug report, the CockroachDB developers realized that a critical predicate normalization rule is missing in the query optimizer. In particular, for an expression that guarantees to yield a non-null result (*e.g.*, comparison between a non-nullable attribute with a non-null value as shown in the above example), it is safe to reduce operations on top of it that still take null value into consideration. With the second query, this rule will remove the IS TRUE check on top of the comparison clause, which leads to a more efficient query execution plan. CockroachDB developers quickly fixed this performance bug given its broad impact on query performance.

EXAMPLE 7. SUB-QUERIES RETURNING SCALAR. The pair of equivalent queries below fetch the same column from the *emp* table with predicates that rely on results of the same subquery. However, when the predicate is false, CockroachDB spends 30× more time to execute the second query compared to the first one. This stems from a performance bug in the implementation of the SCAN operator.

```
/* [First query, 7 milliseconds] */
SELECT sal FROM emp LEFT OUTER JOIN (SELECT job FROM
    bonus LIMIT 1) AS t ON true
WHERE t.job IS NOT DISTINCT FROM 'job';

/* [Second query, 211 milliseconds] */
SELECT sal FROM emp WHERE (SELECT job FROM bonus
    LIMIT 1) IS NOT DISTINCT FROM 'job';
```

With the first query, the DBMS intelligently skips executing the JOIN and SCAN operators after realizing that the predicate in the JOIN operator evaluates to false. However, with the second query, the DBMS ignores the fact that the predicate on the SCAN operator evaluates to false and performs an unnecessary scan of the entire *emp* table. The CockroachDB developers quickly confirmed this performance bug and explained that a potential fix would consist of adding an extra check in the SCAN operator to verify the predicate on top of it. Additionally, they acknowledge that this performance bug belongs to a more important limitation in the query optimizer in that it cannot re-optimize the main query based on the results of the sub-query. They plan to fix this legacy problem in the near future. This would not only fix this performance bug, but also improve the optimizer's ability to handle queries with scalar sub-query results.

EXAMPLE 8. HANDLING AGGREGATE OPERATORS. The pair of equivalent queries below trigger a 2.9× execution time difference on PostgreSQL, which exposes a suboptimal behavior when handling an unnecessary GROUP BY operator.

```
/* [First query, 25 milliseconds] */
SELECT emp_pk FROM emp WHERE emp_pk > 100;

/* [Second query, 72 milliseconds] */
SELECT emp_pk FROM emp WHERE emp_pk > 100 GROUP BY
    emp_pk;
```

Specifically, both queries request the DBMS to fetch the *emp_pk* column based on the same predicate. The second query also appends a GROUP BY operation before returning the final table. Since *emp_pk* is the primary key of the *emp* table, the *emp_pk* column takes unique values, thereby rendering the GROUP BY operation to be unnecessary. However, when PostgreSQL executes the second query, it still performs the GROUP BY operation, which leads to a slower execution time than the first query. The PostgreSQL developers have acknowledged that this performance bug is due to a missing rule in the optimizer to remove unnecessary GROUP BY operations. They plan to add this optimization rule to address this sub-optimal behavior in the future.

DISCUSSION. The empirical results of applying AMOEBA to test CockroachDB and PostgreSQL DBMSs show that AMOEBA is able to effectively detect performance bugs. In particular, we find that semantically equivalent queries do trigger different runtime behaviors in a given DBMS, thereby allowing us to use them as a cross-referencing oracle for finding performance bugs. In addition, we find the presentation of our bug report (*i.e.*, a pair of equivalent queries and their execution plans) provides sufficient information for the DBMS developers. By comparing the execution plans and traces on equivalent queries, the developers are able to quickly pinpoint the root cause of the performance bug.

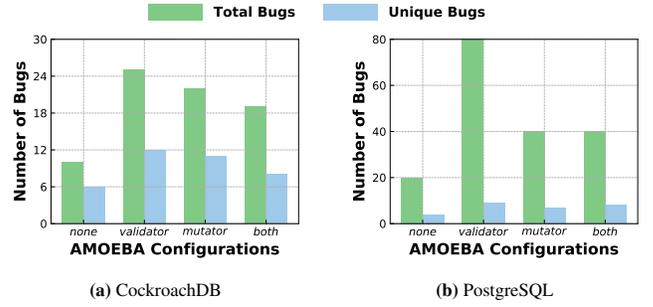

(a) CockroachDB    (b) PostgreSQL

**Figure 6: Efficiency of AMOEBA–** Figures (a) and (b) show the number of performance bugs and unique performance bugs that AMOEBA discovers after five hours in CockroachDB and PostgreSQL, respectively.

> By using the runtime performance of semantically equivalent queries as a cross-referencing oracle, AMOEBA discovers 39 previously-unknown performance bugs spread across different components of two DBMSs.

### 8.4 RQ2 — Efficiency of AMOEBA

We next examine the computational efficiency of AMOEBA in detecting performance bugs. We also investigate whether the feedback mechanisms increase the probability of generating queries that discover performance bugs.

In this experiment, we run AMOEBA on CockroachDB and PostgreSQL across four separate runs using different configurations: (1) AMOEBA$_{none}$ disables both feedback mechanisms, (2) AMOEBA$_{validator}$ enables feedback from VALIDATOR, (3) AMOEBA$_{mutator}$ enables feedback from MUTATOR, and (4) AMOEBA$_{both}$ enables feedback from both MUTATOR and VALIDATOR.

We configure each run to a timeout of 5 hours. Figure 6 presents the results, which include the number of total and unique performance bugs that AMOEBA discovers in each run. We manually map each performance bug report to a corresponding unique bug based on the developers' feedback.

BUG REPORTS AND UNIQUE BUGS. We examine the overall efficiency of AMOEBA by averaging the bug-finding results across four runs. As shown in Figure 6, within 5 hours, AMOEBA generates an average of 19 performance bug reports for CockroachDB (reduces to 9 unique performance bugs). In PostgreSQL, AMOEBA reports an average of 46 performance bugs (reduces to 7 unique performance bugs).

IMPACT OF FEEDBACK MECHANISMS. To better understand the impact of different feedback mechanisms in AMOEBA (*i.e.*, feedback from validator and feedback from mutator), we compare the total and unique performance bugs that AMOEBA discovers across four different runs. For both DBMSs, feedback from validator and mutator increases both the total and unique performance bugs that AMOEBA discovers. On CockroachDB, while AMOEBA$_{none}$ only discovers 10 performance bugs and 6 unique performance bugs, AMOEBA$_{validator}$, AMOEBA$_{mutator}$ and AMOEBA$_{both}$ discover more than 20 performance bugs, which correspond to more than 9 unique performance bugs. On PostgreSQL, while AMOEBA$_{none}$ only discovers 20 performance bugs and 4 unique performance bugs, AMOEBA$_{validator}$, AMOEBA$_{mutator}$ and AMOEBA$_{both}$ discover more than 40 performance bugs, which correspond to more than 7 unique performance bugs.

**RUNTIME PERFORMANCE.** We also count the number of semantically equivalent queries that AMOEBA examines across four runs. On average, AMOEBA generates and examines a pair of semantically equivalent queries per 1.5 and 0.8 seconds for CockroachDB and PostgreSQL, respectively. The overall runtime performance of AMOEBA is dominated by the runtime of the tested DBMS. To further improve the runtime performance of AMOEBA, it is possible to deploy AMOEBA on multiple servers to parallelize the fuzzing loop. We plan to explore this problem in the future.

> AMOEBA detects a number of performance bugs in both CockroachDB and PostgreSQL within a 5 hours run. On both DBMSs, feedback from validator and mutator both have a significant impact on helping AMOEBA discover more performance bugs.

## 8.5 RQ3 — Importance of Mutation Rule

Next, we present an in-depth analysis of mutation rules used by AMOEBA and their importance with respect to discovering performance bugs. We investigate each mutation rule along two dimensions: (1) impact of each rule on query performance, and (2) the frequency of generation of bug-revealing query pairs. We perform this analysis using performance bug reports generated in §8.4. In particular, we examine a dataset of 76 and 182 query pairs that trigger performance bugs in CockroachDB and PostgreSQL, respectively.

**IMPACT ON QUERY PERFORMANCE.** We first examine if a mutation rule speeds up or slows down query execution. If a rule generates a query pair that exhibits a significant runtime performance difference, we measure the speed-up (ratio > 1) or slow-down (ratio < 1). Figure 7 presents the evaluation results in two box plots.

**FREQUENCY OF BUG-REVEALING QUERY PAIRS.** We next compute the frequency with which each mutation rule generate query pairs that trigger performance bugs. To do this, we count the number of times each rule generates such query pairs and normalize this count by the total number of bug-revealing query pairs. Figure 7 presents the evaluation results in two bar plots (the y-axis on the right side denotes the frequency).

The most notable observation in Figure 7 is that not all mutation rules are equally useful in discovering performance bugs. Among 75 mutation rules that AMOEBA uses for generating query pairs, only 39 and 44 rules can generate query pairs that trigger performance bugs in CockroachDB and PostgreSQL, respectively. AMOEBA uses both structure and expression mutation rules for generating these queries.

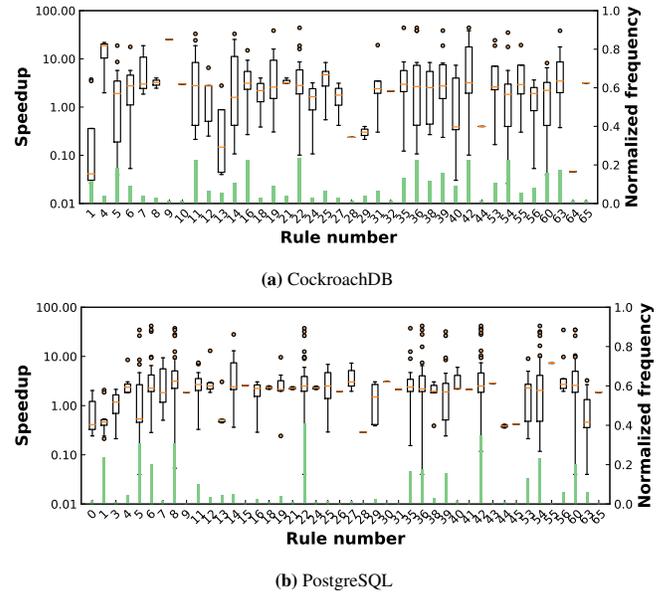

(a) CockroachDB

(b) PostgreSQL

**Figure 7: Impact of Rules on Query Performance and Frequency of Rules That Are Used to Generate Bug-revealing Query Pairs –** The y-axis on the left captures each rule's impact on query performance: a value > 1 indicates the rule is used to generate a query that shows faster runtime performance, while a value < 1 indicates the rule results in a query that shows slower runtime performance. The y-axis on the right captures the normalized frequency of each rule that is used to generate query pairs that trigger performance bugs.

With respect to impact on performance, we find that while some rules always have the same effect on query performance (*i.e.*, either speed-up or slow-down), others exhibit different effects (*i.e.*, both speed-up and slow-down) in different cases. Since AMOEBA seeks to mutate each query by applying a sequence of rules, this patterns result from the compositional effects of mutation rules (§6.1): (1) *contention* between mutation rules: the performance penalty of one rule is damped by the gains from another rule, and vice versa; (2) rules that do not directly affect query performance: while the rule itself does not affect the query performance, it transforms a query to a form that is suitable for mutation by other rules. For instance, rule 22 (match the projected table with its filter predicate) is necessary to trigger other mutation rules that manipulate filter predicates (*e.g.*, rule 16 needs such information to push a filter through a JOIN). So, rule 22 has the highest activation frequency in discovering performance bugs on both DBMSs. Lastly, we summarize the characteristics of mutation rules that generate interesting query pairs, since they may reflect the characteristics of queries that are challenging for DBMS to optimize and execute. For both DBMSs, we find these rules mostly re-arrange or eliminate expensive operators (UNION, GROUP BY, and JOIN) in a given query, while other rules that manipulate the projection and sorting operators (SELECT and ORDER BY) are less likely to generate query pairs that trigger performance bugs. Improving how the DBMS handles these operators in future versions of these DBMSs would enhance their robustness.

> Both structural and expression mutation rules generate queries that reveal performance bugs. We find rules differ in their impact on query performance: while some rules do not directly affect query performance, they are necessary to exercise other rules. In particular, we find that rules that transform expensive operators (*e.g.*, SORT,GROUP BY, and JOIN) are effective in generating queries that trigger performance bugs, which highlights opportunities for improving future versions of the tested DBMSs.

## 8.6 RQ4 — Comparative Analysis

We now present a comparative analysis against two techniques that could also be leveraged to detect performance bugs. The key idea of AMOEBA is to construct semantically equivalent query pairs and use the performance differential within each query pair as a cross-referencing oracle for detecting performance bugs. We now investigate the efficacy of two techniques for obtaining equivalent query pairs for performance bug detection: (1) using SQLancer, an automated DBMS testing technique, to generate random query pairs (§8.6.1). (2) deriving query pairs from the Calcite's test suites, which are carefully crafted by DBMS developers (§8.6.2).

### 8.6.1 Query Pairs from SQLancer

SQLancer is the state-of-art tool for discovering logic bugs [35–37]. A key technique in SQLancer is using TLP to construct equivalent queries based on the observation that any predicate in SQL evaluates to TRUE, FALSE, or NULL. Given a base query, TLP constructs a mutant query that is equivalent to the base query in two steps. First, it partitions the base query into three partition queries, wherein each predicate is constructed based on the value of the base query's predicate. Second, it concatenates these partition queries using the UNION or UNION ALL operator based on the semantics of the base query. Taking the query pair shown below as an example, the boolean predicate t0.c0 = t1.c0 from the base query is compared against TRUE, FALSE, or NULL in each of the partition queries. TLP constructs the mutant query by concatenating these partition queries with the UNION ALL operator.

**EXAMPLE 9. QUERY GENERATED BY TLP.**

```
/* [Base query] */
SELECT * FROM t0, t1 WHERE t0.c0 = t1.c0;
/* [Mutant query] */
SELECT * FROM t0, t1 WHERE t0.c0 = t1.c0
UNION ALL
SELECT * FROM t0, t1 WHERE NOT (t0.c0 = t1.c0)
UNION ALL
SELECT * FROM t0, t1 WHERE (t0.c0 = t1. c0) IS NULL;
```

### 8.6.2 Manually-Crafted Query Pairs

We derive a set of semantically equivalent query pairs from Calcite's test suite [17, 46]. These tests are manually crafted to ensure the correctness of query transformation rules in Calcite. Each test contains a SQL query and a set of transformation rules under test. Each test case transforms the query's logic query plan tree using the rules and examines whether the resulting logic query plan tree is expected. We derive pairs of semantically equivalent queries from these tests: we use the input query as the base query and the transformed query as the mutant query.

This human-intensive baseline is challenging because: (1) they cover a wide range of SQL operators, and (2) they use the same set of transformation rules as AMOEBA does that are effective at discovering performance bugs (§8.3).

| Base Query | Mutant Query | Performance Bugs Found | |
|---|---|---|---|
| | | Cockroach | PostgreSQL |
| SQLancer | SQLancer | 0 | 0 |
| Calcite Test | Calcite Test | 4 | 4 |
| Calcite Test | AMOEBA | 4 | 6 |
| AMOEBA | AMOEBA | 25 | 14 |

**Table 4: Comparative Analysis of AMOEBA–** The results include the number of performance bugs that each set of query pairs is able to uncover with the SCOTT database. We provide the total number of bugs discovered by AMOEBA in the fourth row for comparison.

### 8.6.3 Results

As shown in Table 4, we compare AMOEBA against three baselines: (1) We use SQLancer to randomly generate 2000 pairs of equivalent queries. (2) We derive 373 pairs of equivalent queries from the Calcite test suite. (3) We acquire another 373 pairs of equivalent queries by using AMOEBA's MUTATOR to mutate base queries derived from the Calcite test suite. Then, we run each query pair on a given target database and measure the number of performance bugs that they reveal.

The most notable observation is that AMOEBA finds significantly more performance bugs than alternatives. Specifically, AMOEBA discovers 25 and 14 performance bugs in CockroachDB and PostgreSQL, respectively. Manually-crafted query pairs derived from the Calcite test suite discover only 4 and 6 performance bugs in each DBMS, respectively. This illustrates the utility of automating the query transformation process. The query pairs generated by the SQLancer reveal zero performance bugs. We next analyze the factors that contribute to the efficacy of AMOEBA.

**MUTATION RULES AND ALGORITHM.** SQLancer differs from AMOEBA in that it mutates queries using TLP. We find that TLP is not useful for detecting performance bugs. By design, the TLP query is more complex than the corresponding base query, which inevitably leads to a higher execution time. Among 2000 query pairs generated by this technique, we discover that 16 and 12 pairs of equivalent queries exhibit a significant difference in execution time for PostgreSQL and CockroachDB, respectively. Among these queries, we find that the TLP query always takes more time to execute compared to the base query (with an average slow-down of 17 ×). The root cause of these performance differences is that the TLP query forces the DBMS to perform additional operations (*i.e.*, fetching the tuples for each partition query and then combining those results together). Given this inherent overhead, TLP is not able to find a variety of performance bugs, which are discovered by AMOEBA and Calcite's test suite.

While Calcite and AMOEBA share the same set of mutation rules, they differ in how they leverage mutation rules. Each Calcite test transforms the base query using a small set of mutation rules with a specific firing order. On the other hand, AMOEBA mutates the same base query with different combination of all the rules. As shown in the second and third rows of Table 4, AMOEBA's mutation strategy is more likely to discover performance bugs than the manual efforts. With PostgreSQL, AMOEBA's mutation algorithm discovers two more performance bugs than Calcite. By design, for each test case in Calcite, the base query is only transformed by a specific

Testing DBMS Performance with Mutations

|  | Base Query | Function | Keyword | Join Type | Data Type |
|---|---|---|---|---|---|
| CockroachDB | CALCITE | 15 (8) | 17 | 3 | 4 |
|  | AMOEBA | 7 | 18 (1) | 3 | 4 |
| PostgreSQL | CALCITE | 16 (9) | 18 | 3 | 4 |
|  | AMOEBA | 7 | 18 | 3 | 4 |

**Table 5: Support of SQL Clauses and Types –** Number of SQL clauses and types supported for CockroachDB and PostgreSQL DBMSs. Number inside () denotes the number of operators supported only by one system.

sequence of rules. Such a mutation procedure cannot generate a variety of queries that are equivalent to the same base query, which lowers the probability of acquiring appropriate query pairs to discover performance bugs. On the other hand, AMOEBA's automated mutation strategy exploits all available query transformation rules and their compositional effects to generate equivalent query pairs, which increases chances of discovering performance bugs (§6.2).

As shown in Table 4, base queries derived from the Calcite test suite are not sufficient to discover all performance bugs that AMOEBA discovered. Calcite only uncovers 4 of 25 performance bugs in CockroachDB and 6 of 14 performance bugs in PostgreSQL, respectively.

## 8.7 RQ5 — Analysis of Base Queries

To understand why AMOEBA outperforms the manually crafted test suite, we next compare their base queries. For this analysis, we construct a dataset of base queries: (1) We use AMOEBA to randomly generate 2000 base queries for CockroachDB and PostgreSQL, and (2) We extract 373 base queries from Calcite.

**SINGLE CLAUSE ANALYSIS.** We first examine the SQL clauses and type coverage for both AMOEBA and Calcite. We find that AMOEBA and Calcite tests almost cover the same set of SQL types and operators (*e.g.*, join operators and common keywords). Calcite even covers more function names than AMOEBA does. Since Calcite's base queries are less effective than AMOEBA's base queries at discovering performance bugs, we infer that higher single-clause coverage does not increase the likelihood of generating a base query that triggers a performance bug. So, we next investigate the importance of combinations of different SQL clauses.

**TWO-CLAUSE COMBINATION ANALYSIS.** We examine two-clause combination coverage for both AMOEBA and Calcite (*e.g.*, whether a base query can have both GROUP BY and LEFT JOIN). To do this, we construct the co-occurrence matrix using the base query dataset and SQL clauses presented in Table 5 [1]. For each base query dataset, we first count the frequency of having two SQL clauses in the same query, normalize the count by the number of total queries, and then plot the result in a heatmap. Additionally, we identify clause pair combinations within base queries that trigger performance bugs (§8.4), and highlight those combinations that are only found by AMOEBA using ∗ marker in the heatmap shown in Figure 8. We only present the heatmap for CockroachDB (similar trend in PostgreSQL). Since AMOEBA-generated base queries lead to discover more performance bugs, we believe these highlighted combinations represent *interesting* two-clause combinations that are more likely to trigger performance bugs. The results show that, for both DBMSs, base queries generated by AMOEBA explore significantly larger space of two-clause combinations than Calcite tests. While base queries from Calcite tests cover 103 and 106 clause pair combinations for CockroachDB and PostgreSQL, base queries from AMOEBA cover 208 and 200 combinations for the two DBMSs, respectively.

[1] We merge all function names into one FUNC clause to save space

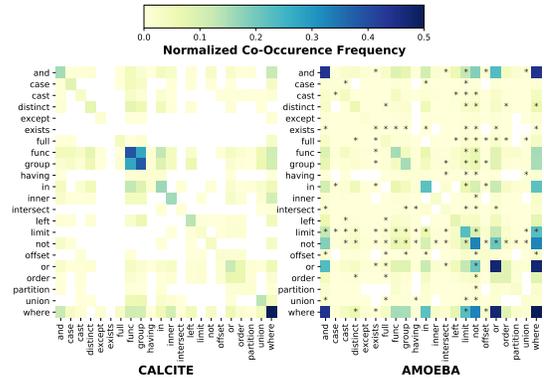

**Figure 8: Two-Clause Combinations –** AMOEBA-generated base queries cover significantly more clause pair combinations than those covered by Calcite tests. As a result, AMOEBA covers certain *interesting* patterns (highlighted with the ∗ marker) that uncover performance bugs.

**SUGGESTIONS FOR DBMS DEVELOPERS.** The results also show that base queries from Calcite tests miss a significant amount of *interesting* clause pair combinations that lead to discover performance bugs. Specifically, they miss 58 and 52 clause pair combinations for CockroachDB and PostgreSQL, respectively. We summarize the characteristics of these clause pairs that the Calcite fails to cover, since they may reflect query patterns that are challenging for DBMS to optimize but are neglected by manual testing efforts. For both DBMSs, we find: (1) filter clauses (*i.e.*, WHERE and HAVING) are important for discovering performance bugs. When combined with expensive operators, such as JOIN, GROUP BY, and UNION, their placements (before or after these operators) play a significant role in deciding query execution time. (2) The LIMIT clause challenges the optimizer. Since LIMIT asks for a smaller set of results, an optimal plan should either scan a partial table or terminate expensive operations early while still fetching the correct result sets. Improving how the DBMS handles these operations and their combinations would enhance its performance robustness.

> AMOEBA outperforms SQLancer and Calcite for three reasons: (1) AMOEBA differs from SQLancer in that it leverages a variety of optimization rules to mutate the base query, which are more likely to trigger different runtime behaviors that expose performance bugs; (2) Compared with Calcite, AMOEBA broadly explores the equivalent query space by leveraging all available rules and their compositional effects; (3) Compared with Calcite, AMOEBA covers more interesting clause pair combinations that are challenging for the DBMS to optimize and execute.

## 9 Limitations and Future Work

We now discuss the limitations of AMOEBA and present ideas for tackling them in the future.

**DIFFERENTIAL TESTING.** Since AMOEBA is based on a widely-used query optimization framework (*i.e.*, Calcite), it inherits the limitations of differential testing. First, it only focuses a widely-supported subset of SQL operators and functions. Fortunately, since Calcite is an extensible framework, it is feasible to add support for additional SQL features. Second, it assumes that a given SQL feature has the same semantics across all the DBMSs under the test. In particular, AMOEBA performs semantics-preserving transformations based on the widely-used semantics of common SQL features.

**QUERY OPTIMIZATION VS EXECUTION TIME.** Finding the optimal execution plan is an NP-hard problem [23]. So, the DBMS may choose to settle for a sub-optimal plan. Given this tradeoff between optimization and execution time, it is challenging for developers to strike a perfect balance between them. However, the performance bugs that AMOEBA uncovers provide challenging test-cases that encourage developers to carefully examine this tradeoff.

## 10 Related Work

In this section, we present prior work on testing DBMSs with an emphasis on DBMS performance.

**FUZZING DBMSS.** Given the large state space of possible SQL queries, fuzzing has been applied to find crash bugs and security vulnerabilities in DBMSs [2–4]. Researchers have improved the efficacy of the fuzzing loop by taking the feedback from the tested DBMS into consideration [12, 45]. While AMOEBA is also a fuzzing tool equipped with a feedback mechanism, it differs from prior work in that it focuses on generating semantically equivalent query pairs that trigger different runtime behaviors.

**LOGIC BUGS.** To circumvent the oracle availability problem associated with automated testing, researchers have applied *differential* and *metamorphic* testing techniques for discovering logic bugs in DBMSs [16, 29]. RAGS discovers logic bugs by executing the same query on different DBMSs and comparing the results [40]. Waas *et al.* propose a framework for validating the query optimizer by executing alternative execution plans for the input query and comparing their results [41]. These techniques are not tailored for discovering performance bugs.

SQLancer is the state-of-the-art tool for discovering logic bugs in DBMS using metamorphic testing [35–37]. The key idea behind SQLancer is to construct a metamorphic relation that is used to generate a cross-referencing oracle for detecting logic bugs. AMOEBA is similar to SQLancer in that it uses semantics-preserving query mutation rules to establish a metamorphic relation for discovering performance bugs. As discussed in §8.6, while the metamorphic relation proposed in SQLancer is effective in detecting logic bugs in DBMS, it is not suitable for discovering performance bugs. To the best of our knowledge, AMOEBA is the first technique that uses metamorphic testing to discover performance bugs in DBMS.

**PERFORMANCE TESTING.** Researchers have presented techniques for finding performance bugs by executing the DBMS on a predefined workloads and comparing their behavior against performance baselines [24, 34, 43, 44]. These techniques detect performance regressions stemming from DBMS upgrades and configuration changes. AMOEBA differs from these approaches in that it does not require a pre-defined baseline for finding performance bugs. Instead, it leverages the tested DBMS's runtime behaviors on equivalent queries as a cross-referencing oracle.

**OPTIMIZER TESTING.** Researchers have proposed techniques for testing the query optimizer's ability to find the best execution plan [19, 22]. Li *et al.* propose a benchmark for assessing the efficiency of a query optimizer (*i.e.*, optimization time) [28]. Leis *et al.* investigate the impact of the components of the query optimizer on runtime performance [27]. These efforts are geared towards quantifying the quality of an optimizer. Another line of research focuses on developing frameworks for testing the correctness of query transformation rules in the query optimizer [18, 41]. These efforts require an in-depth knowledge about the tested query optimizer. AMOEBA complements these efforts by taking a black-box approach and facilitates more extensive testing of optimizers.

## 11 Conclusion

In this paper, we presented a novel and effective approach for detecting performance bugs in DBMSs. Our key idea is to construct a semantically equivalent query pair and then compare their runtime performance. If the two queries show a significant difference in their execution time, then the likely root cause is a performance bug in the tested DBMS. We proposed a novel set of structure and expression mutation rules for constructing query pairs that are likely to uncover performance bugs. We implemented this technique in a tool called AMOEBA. We introduced feedback mechanisms for improving the efficacy and computational efficiency of the tool. We evaluated AMOEBA on two widely-used DBMSs and discovered 39 previously-unknown performance bugs. Our empirical analysis of the factors that help AMOEBA discover more performance bugs than alternative techniques highlights opportunities for improving future versions of these DBMSs.